\pdfoutput=1
\documentclass[twocolumn,aps]{revtex4}
\usepackage{graphicx}
\usepackage{amsmath}
\begin{document}

\def\beq{\begin{equation}}
\def\eeq{\end{equation}}
\def\sss{\scriptscriptstyle\rm}
\def\x{_{\sss X}}
\def\c{_{\sss C}}
\def\s{_{\sss S}}
\def\xc{_{\sss XC}}
\def\dc{_{\sss DC}}
\def\ext{_{\rm ext}}
\def\ee{_{\rm ee}}
\def\sint{ {\int d^3 r \,}}
\def\br{{\mathbf r}}

\title{Adiabatic Connection for Strictly-Correlated Electrons}
\author{Zhen-Fei Liu and Kieron Burke}
\affiliation{Department of Chemistry, University of California, Irvine, California, 92697-2025, USA}
\date{\today}

\begin{abstract}
Modern density functional theory (DFT) calculations employ the Kohn-Sham (KS) system of non-interacting electrons as a reference, with all complications buried in the exchange-correlation energy ($E\xc$). The adiabatic connection formula gives an exact expression for $E\xc$. We consider DFT calculations that instead employ a reference of strictly-correlated electrons. We define a ``decorrelation energy'' that relates this reference to the real system, and derive the corresponding adiabatic connection formula. We illustrate this theory in three situations, namely the uniform electron gas, Hooke's atom, and the stretched hydrogen molecule. The adiabatic connection for strictly-correlated electrons provides an alternative perspective for understanding density functional theory and constructing approximate functionals.
\end{abstract}

\maketitle

\section{Introduction}

For most modern calculations using density functional theory (DFT) \cite{Kb99}, the accuracy of results depends only on approximations to the exchange-correlation functional, $E\xc[n]$. An exact expression for $E\xc[n]$ is given by the adiabatic connection formula \cite{LP75, GL76}, in which $E\xc$ is expressed as an integral over a coupling constant $\lambda$, which connects the reference (Kohn-Sham system, $\lambda=0$) and the real physical interacting system (ground state, $\lambda=1$), keeping the density $n(\br)$ fixed. Study of the adiabatic connection integral has proven very useful for understanding approximate (hybrid) functionals \cite{MCY06, PMTT08}, and is an ongoing area of research \cite{BEP97, FTB00, MTB03}.

Almost all modern DFT calculations employ the Kohn-Sham (KS) system \cite{KS65} as a reference. The KS system is defined as the unique fictitious system that has the same density as the real system, but consists of non-interacting electrons. The great practicality of KS DFT is due to the relative ease with which the non-interacting equations can be solved, with relatively crude approximations, giving KS DFT a useful balance between speed and accuracy.

However, DFT calculations could also be based on another fictitious system which is known as the strictly-correlated (SC) system \cite{SPL99}. The strictly-correlated system has the same density as the real system (as does the KS system), but the Hamiltonian consists of electron repulsion and external potential energy terms only. In recent years, the pioneering work of Seidl and others \cite{S99, SPL99, SPK00, SPKb00, SGS07, S07, GSS08, GVS09} has led to substantial progress in solving this problem exactly and efficiently. The strictly-correlated electrons (SCE) ansatz \cite {SGS07,S07,GSS08,GVS09} has been shown to yield the density and energy of this system, going beyond the earlier point-charge-plus-continuum (PC) model \cite{SPK00,SPKb00}. They have achieved great success in calculating spherical symmetric systems with arbitrary number of electrons \cite{S07}.

In this article, we look to the future and assume that the strictly-correlated limit of \emph{any} system can be calculated with less difficulty than the original interacting problem, and all our successive work is developed based on this reference. We derive a new version of the adiabatic connection formalism, which connects the strictly-correlated system (fully interacting) and the physical system. We also introduce a new coupling constant $\mu$, and a ``decorrelation energy'' $E\dc$, the counterpart of $E\xc$ in KS DFT, which must be evaluated to extract the true ground-state energy from the calculation of the strictly-correlated system. We argue that, as long as the strictly-correlated system can be solved easily (just as the KS case), one can develop another version of DFT based on this system, a version that is better-suited to strongly localized electrons.

Throughout this paper, we use atomic units ($e^2=\hbar=\mu=1$), which means that if not particularly mentioned, all energies are in Hartrees, and all lengths are in Bohr radii, etc.

\section{Theory}

In this section, we introduce the alternative adiabatic connection formula, and relate its quantities to more familiar ones.  All results here are formally exact.

\subsection{Kohn-Sham Adiabatic Connection}

In KS DFT, the total energy for the interacting ground-state is expressed as:
\begin{equation}
E[n] = T\s[n] + \sint v\ext(\mathbf{r}) \, n(\mathbf{r}) + U[n] + E\xc[n].
\label{eings}
\end{equation}
In this equation, $T\s$ is non-interacting kinetic energy of KS orbitals $\{ \varphi_i \}$ that are eigenfunctions of the non-interacting KS equation, $v\ext(\br)$ is external potential (nuclear attraction in the case of atoms and molecules), $U$ is the Hartree energy defined as the ``classical'' Coulomb repulsion between two electron clouds, and $E\xc$ is the exchange-correlation energy \cite{primer}. The adiabatic connection integral \cite{LP75,GL76} then gives an exact expression for $E\xc$:
\begin{equation}
E\xc [n] = \int_0^1 d\lambda \, W_\lambda [n],
\label{normalac}
\end{equation}
where one can show that $W_\lambda= < \Psi^\lambda | \hat{V}\ee | \Psi^\lambda > - U$, in which $\Psi^{\lambda}$ is the wavefunction that minimizes $\hat{T}+\lambda \hat{V}\ee$ but has the same density as the real ground-state system \cite{Y87}. At $\lambda=0$, one recovers the KS system, and at $\lambda=1$, one recovers the real interacting system. In this way, one connects the KS system with the real interacting system by changing $\lambda$ from $0$ to $1$.

\begin{figure}[h]
\begin{center}
\includegraphics[width=3.5in]{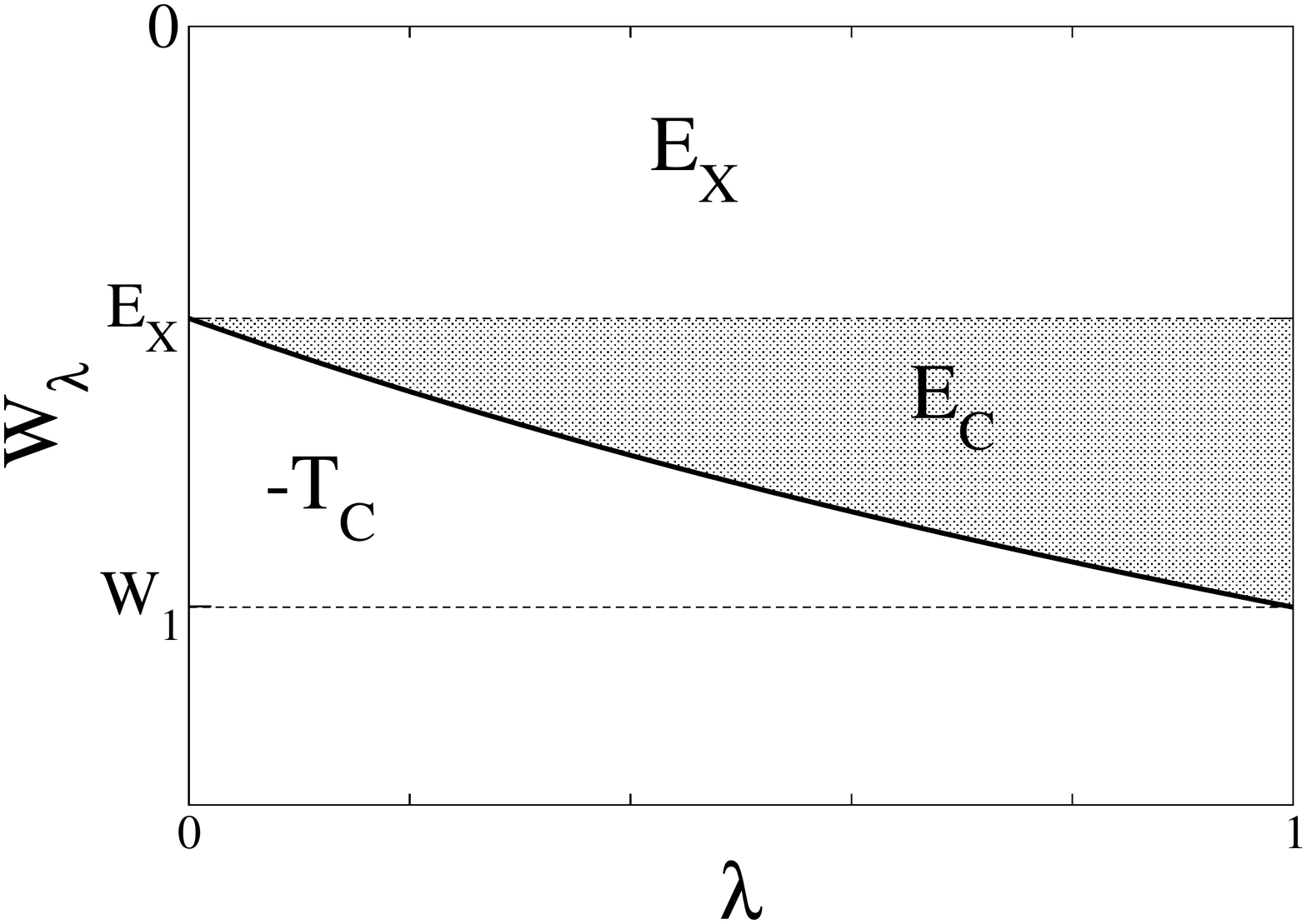}
\includegraphics[width=3.5in]{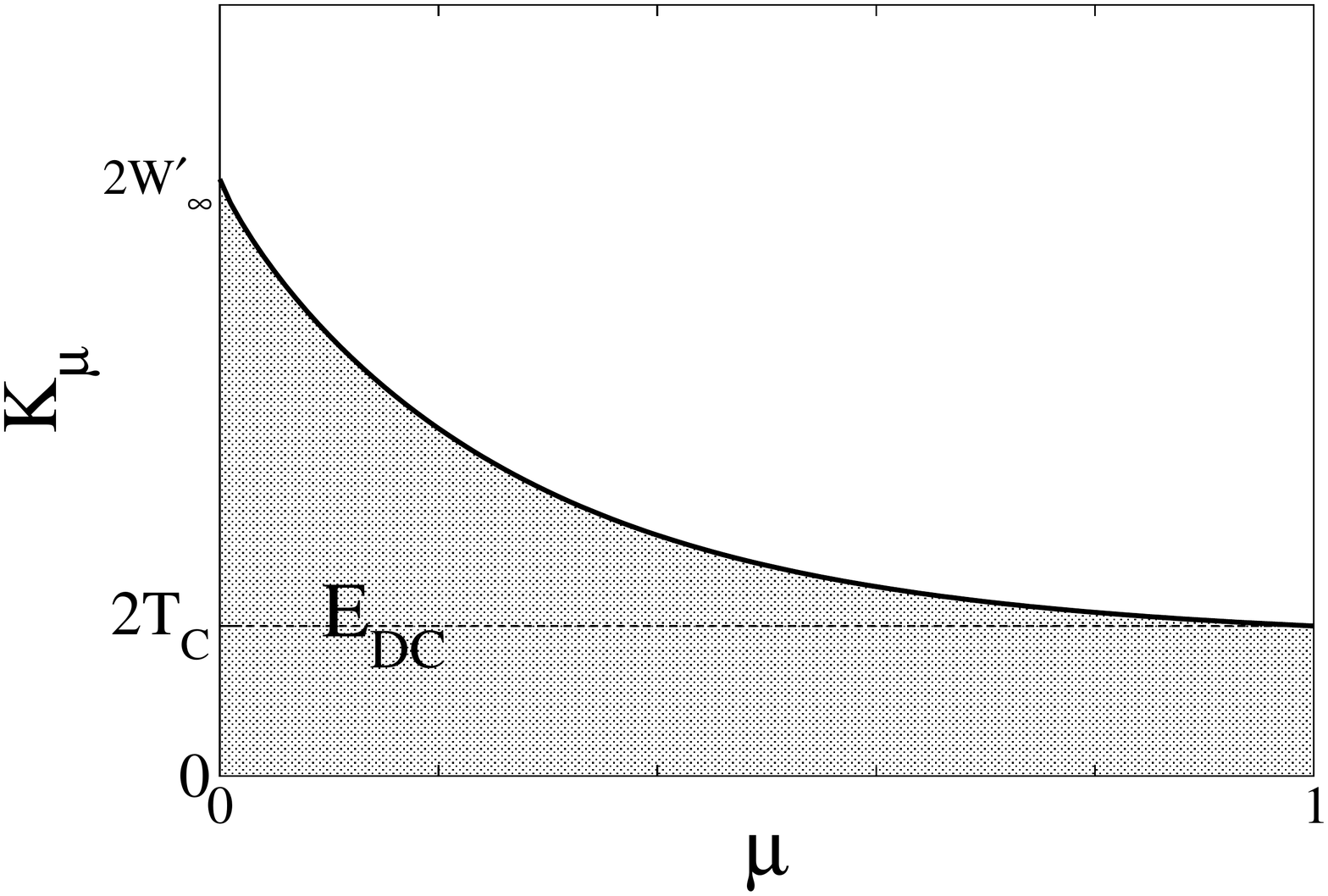}
\caption{Traditional (upper panel) and the new (lower panel) adiabatic connection curves.}
\label{cartoons}
\end{center}
\end{figure}

A cartoon of $W_\lambda$ versus $\lambda$ is shown in the upper panel of Fig. \ref{cartoons}. By definition, we have $W_0=E\x$ and the area under the curve is $E\xc$. We can also identify the kinetic correlation energy $T\c=E\xc-W_1$ in this graph.

\subsection{Strictly-Correlated Reference System}

The KS wavefunction can be defined as the wavefunction that minimizes the kinetic energy alone, but whose density matches that of the interacting system.  Analogously, the SC wavefunction is found by minimizing the electron-electron repulsion energy alone, subject to reproducing the exact density.  In practice, there might be multiple degeneracies, so it is best defined in the limit as the kinetic energy goes to zero.

Then, using the strictly-correlated (SC) system as the reference, the energy of the true interacting ground state is:
\begin{equation}
E[n] = U_{\rm sc}[n] + \sint v\ext(\mathbf{r}) \, n(\mathbf{r}) + T\s [n] + E\dc[n],
\label{eingswsc}
\end{equation}
where $U_{\rm sc}=<\Psi^\infty |\hat{V}\ee|\Psi^\infty>$. In KS DFT quantities, $U_{\rm sc}=U+W_\infty$ [see Eq. (\ref{normalac})].

Just as we separate the Hartree energy from the total energy in KS DFT [Eq. (\ref{eings})], here in Eq. (\ref{eingswsc}) we separate $T\s[n]$ from the total energy in SC DFT. There are a variety of algorithms that can be used to extract this quantity accurately for any given density, effectively by inverting the KS equations \cite{PVW03}. We label the remainder as the  ``decorrelation energy'', $E\dc[n]$. The reason we call it ``decorrelation energy'' is that, if we consider the electrons in the reference system ``strictly correlated'', with energy $U_{\rm sc}[n]$, the electrons in the real system are \emph{less} correlated than in the reference system. We will see the physical meaning explicitly very soon.

So far, we have defined our reference, and next we deduce an exact expression for the newly-defined ``decorrelation energy'' $E\dc[n]$ with the adiabatic connection formalism, just as one does for $E\xc[n]$ \cite{SPL99, SPK00} in the KS DFT [Eq.( \ref{normalac})].

\subsection{Strictly-Correlated Adiabatic Connection Formula}

We denote $\Psi^{\mu}$ as the wavefunction minimizing $\hat{H}^{\mu} = \mu^2 \hat{T} + \hat{V}\ee + \hat{V}\ext^{\mu}$ with density $n(\br)$. For $\mu=0$, we recover the strictly-correlated system, and for $\mu=1$, we recover the real system.  For each value of $\mu$, there is a corresponding unique external potential yielding the correct density, $v\ext^{\mu}(\br)$. So we have:
\begin{equation}
E^{\mu}
=\langle \Psi^\mu|\, \mu^2 \hat{T} + \hat{V}\ee + \hat{V}\ext^{\mu} | \Psi^{\mu} \rangle
\label{minihmu}
\end{equation}
Using Hellmann-Feynman theorem \cite{LP85}, we have:
\begin{equation}
\frac{dE^{\mu}}{d\mu}=\left< \Psi^{\mu} \left| \frac{d\hat{H}^{\mu}}{d\mu} \right| \Psi^{\mu} \right> = \left< \Psi^{\mu} \left| 2\mu \hat{T} + \frac{d\hat{V}\ext^{\mu}}{d\mu} \right| \Psi^{\mu} \right>.
\label{hft}
\end{equation}
Integrating and cancelling the external potential terms at both sides, we recognize the left hand side is just $T\s[n]+E\dc[n]$. Thus:
\begin{equation}
E\dc[n] = \int_0^1 d\mu \, 2\mu \left< \Psi^{\mu} \left| \hat{T} \right| \Psi^{\mu} \right> - T\s[n].
\label{ac}
\end{equation}
This is our adiabatic connection formula for strictly-correlated electrons. Finally, since $T\c[n] = T[n] - T\s[n]$, and $T\s[n]$ is independent of $\mu$:
\begin{equation}
E\dc[n] = 
\int_0^1 d\mu \, K_\mu[n],
\label{updnac}
\end{equation}
where
\begin{equation}
K_\mu[n] = 2\mu\, T\c^\mu[n].
\label{Kmudef}
\end{equation}
This is the SC doppelganger of Eq. (\ref{normalac}). We plot a cartoon of the integrand $K_\mu$ vs. $\mu$ in the lower panel of Fig. \ref{cartoons}, identifying the area below the curve as $E\dc$, and noting that $K_1=2T\c$.

\subsection{Relation to Kohn-Sham DFT}

From a formal viewpoint, what we derived here is not new, but simply another way to describe the real interacting system. Thus we can relate all quantities defined here, such as $E\dc[n]$ and $K_{\mu}[n]$, to quantities defined in the traditional KS DFT. Since both Eq. (\ref{eingswsc}) and Eq. (\ref{eings}) are exact for the real system, and if we use the expression of $U_{\rm sc}[n]$ in KS language [see discussion below Eq. (\ref{eingswsc})], we find:
\begin{equation}
E\dc[n] = E\xc[n] - W_\infty[n].
\label{edcexc}
\end{equation}
Thus $E\dc[n]$ defined in our theory is just the difference between the usual exchange-correlation energy of the real system, $E\xc[n]$, and the potential contribution to the exchange-correlation energy of the strictly-correlated system, $W_\infty[n]$.

We can also deduce an expression for $K_\mu$ in terms of $W_\lambda$.  Since $\Psi^{\mu}$ minimizes $\hat{H}^{\mu}=\mu^2 \hat{T}+\hat{V}\ee$ while yielding $n(\br)$, and $\Psi^{\lambda}$ minimizes $\hat{T}+\lambda \hat{V}\ee$, we deduce $\Psi^{1/ \mu^2}=\Psi^{\lambda}$. Now, from the scaling properties of KS DFT \cite{L91}, we know:
\begin{equation}
T\c^{\lambda}=E\c^{\lambda}-\lambda \frac{dE\c^{\lambda}}{d\lambda}.
\label{inttc}
\end{equation}
If we write $E\c^{\lambda}=T\c^{\lambda}+U\c^{\lambda}$, i.e., $U\c^\lambda$ is the potential contribtion to $E\c^\lambda$, $U\c^\lambda=\lambda (W_\lambda - E\x)$, we have:
\begin{equation}
\frac{dT\c^{\lambda}}{d\lambda}=\frac{U\c^{\lambda}}{\lambda}-\frac{dU\c^{\lambda}}{d\lambda}.
\label{difftc}
\end{equation}
Integrating over $\lambda$ from 0 to $1/ \mu^2$, and using the definition of $W_\lambda$ in Eq. (\ref{normalac}) and that $E\x^{\lambda}=\lambda E\x$ by scaling \cite{L91}, we can express $T\c^{\mu}[n]$ in terms of $W_\lambda[n]$, we find:
\begin{equation}
K_\mu[n] = 2\mu \int_0^{1/ \mu^2} d\lambda \left( W_\lambda[n] - W_{1/ \mu^2}[n] \right).
\label{wl4kmu}
\end{equation}
From this relation, we can generate the new adiabatic connection curve, as long as we know the integrand of the KS adiabatic connection, i.e.  $W_\lambda[n]$ for $\lambda=1$ to $\infty$.

\subsection{Exact conditions}

Many of the well-established exact conditions on the correlation energy can be translated and applied to the decorrelation energy. In particular, the simple relations between scaling the density and altering the coupling constant all apply, i.e.,
\beq
E\c^\lambda[n]=\lambda^2\, E\c[n_{1/\lambda}],
\eeq
where $n_\gamma(\br) = \gamma^3\, n(\gamma \br)$.  Thus, in terms of scaling:
\beq
K_\mu [n] = \frac{2}{\mu^3} T\c[n_{\mu^2}].
\eeq
Note that, as $\mu \to \infty$, $K_\mu \to 0$, while $K_{\mu=0}=2W_{\infty}'$, where $W_{\infty}'$ is defined in the expansion of $W_\lambda$ as $\lambda \to \infty$ \cite{SPK00}:
\begin{equation}
W_\infty'=\lim_{\lambda \to \infty} \sqrt{\lambda} \left( W_\lambda-W_\infty \right).
\label{watinfty}
\end{equation}
Thus the SC energy is found from solving the strictly-correlated system, while $K_{\mu=0}$ is determined by the zero-point oscillations around that solution. Both are currently calculable for spherical systems \cite{S07,GVS09}.

The most general property known about the correlation energy \cite{L91} is that, under scaling toward lower densities, it must become more negative. In turn, this implies that $W_\lambda$ is never positive. Using the definition of $T\c^\mu$ and changing variable $\lambda=1/\mu^2$ in Eq. (\ref{difftc}), we find:
\beq
\frac{dT\c^\mu}{d\mu}=\frac{2}{\mu^5}\frac{dW_\lambda}{d\lambda}<0,
\label{tcmul0}
\eeq
then using $K_\mu=2\mu T\c^\mu$ and the fact that $K_\mu>0$, we find:
\beq
\frac{d}{d\mu} \ln K_\mu <0.
\label{inequality}
\eeq
Also, because $T\c^\mu>0$, so $K_\mu=2\mu T\c^\mu>0$, and $E\dc$, as an integration of $K_\mu$, is always positive.

Based on these properties of $K_\mu$, a crude approximation to $K_\mu$ can be a simple exponential parametrization, using $K_0$ and the derivative of $\ln K$ at $\mu=0$ as inputs:
\beq
K=K_0\,e^{-\gamma \mu}, \hspace{0.5in} \gamma=-\left. \frac{d}{d\mu} \ln K \right|_0.
\label{simpleunif}
\eeq

\section{Illustrations}
In this section, we illustrate the theory developed above on three different systems, to show how $K_\mu$ behaves for very different systems, and where the adiabatic connection formula might be most usefully approximated.

\subsection{Uniform Electron Gas}

For a uniform electron gas, we assume we know the correlation energy per particle, $\epsilon\c$, accurately as a function of $r_{s}=(3/4\pi n)^{1/3}$. In order to apply Eq. (\ref{wl4kmu}) to calculate $K_\mu[n]$, we use $\epsilon \c^{\lambda} (r_s) = \lambda ^2 \epsilon \c (\lambda r_s)$ \cite{L91}. Substituting into Eq. (\ref{inttc}), changing variables $\lambda=1/ \mu^2$, and using $K_\mu=2 \mu T\c^{\mu}=N \kappa_\mu$, with $N$ the number of particles, we find:
\beq
\kappa_\mu^{\rm unif}  = -\frac{2}{\mu^3} \frac{d}{dr_s} \left. \left( r_s \epsilon\c(r_s) \right) \right| _{r_s/\mu^2}.
\label{unigaskmu}
\eeq
Using Eq. (\ref{edcexc}) and the definition of $W_\infty$, we find:
\beq
\epsilon\dc^{\rm unif}=\epsilon\c+\frac{d_0}{r_s},
\label{edcunif}
\eeq
where $d_0$ is defined below and $d_0=0.433521$. In the large-$r_s$ limit or the low-density expansion \cite{PW92}: 
\begin{equation}
\epsilon \c (r_s) = - \frac{d_0}{r_s} + \frac{d_1}{r_s^{3/2}} + \frac{d_2}{r_s^2}+\cdots
\label{unigasec}
\end{equation}
where $d_2=-3.66151$ from data of Ref. \cite{PW92}. Substituting this expansion into Eq. (\ref{unigaskmu}), we find:
\beq
\kappa_\mu^{\rm unif} = \frac{d_1}{r_s^{3/2}}+ 2 \mu \frac{d_2}{r_s^2}+\cdots \hspace{0.5in} \mbox{as $\mu \to 0$}.
\label{unigaskmuexpn}
\eeq
Thus $\kappa_\mu$ is expected to have a well-behaved expansion in powers of
$\mu$ for small $\mu$.

\begin{figure}[h]
\begin{center}
\includegraphics[width=3.5in]{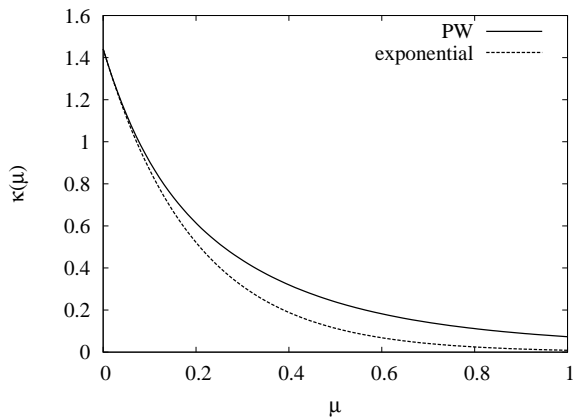}
\caption{Exact adiabatic connection curve $\kappa_{\mu}$ for uniform electron gas ($r_s=1$) and a simple exponential parametrization.}
\label{unigasplot}
\end{center}
\end{figure}

Using Perdew and Wang's \cite{PW92} parametrization of the correlation energy of the uniform gas, we plot $\kappa_\mu$ vs. $\mu$ for $r_s=1$ in Fig. \ref{unigasplot}, and find $\epsilon\dc=0.374$ at $r_s=1$.

Using the exact curve for $r_s=1$ in the simple exponential parametrization [Eq. (\ref{simpleunif})], we find $\kappa_0=1.44073$ and $\gamma=5.0826$. We plot the exponential parametrization in Fig. \ref{unigasplot} and we can see that it decays much faster than the exact curve, producing a $\epsilon\dc$ that is too small by about 25\%, which means about 150\% larger in $|\epsilon\c|$ [see Eq. (\ref{edcexc})].

\begin{figure}[h]
\begin{center}
\includegraphics[width=3.5in]{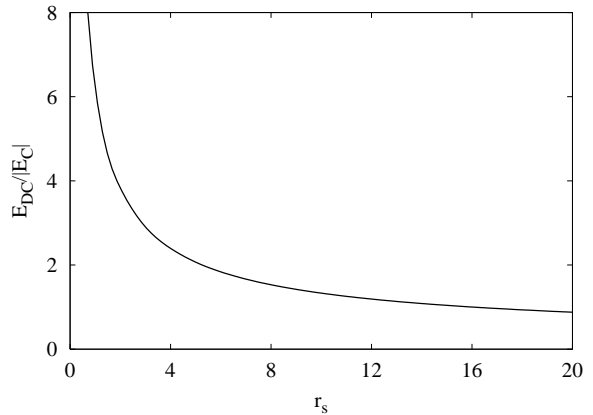}
\caption{$\epsilon\dc / |\epsilon\c|$ for different $r_s$ for uniform electron gas.}
\label{diffrs}
\end{center}
\end{figure}

We calculated $\epsilon\dc / |\epsilon\c|$ for different values of $r_s$ and plot the curve in Fig. \ref{diffrs}. At small $r_s$, $\epsilon\dc \gg |\epsilon\c|$, which suggests that the KS reference system is a better starting point, as a smaller contribution to the energy needs to be approximated. At large $r_s$, $|\epsilon\c| > \epsilon\dc$ so $\epsilon\dc$ is a smaller quantity and may be better approximated. Under such circumstances, the strictly-correlated system might serve as a better reference. For the uniform gas, the switch-over occurs at about $r_s=16$, which is at densities much lower than those relevant to typical processes of chemical and physical interest.  However, as we show below, for systems with static correlation, this regime can occur much more easily.

\subsection{Hooke's Atom}
As we pointed out, as long as we have an approximate $W_\lambda[n]$ for $\lambda$ between 1 and $\infty$, we can substitute it into Eq. (\ref{wl4kmu}) to get the new adiabatic connection formula for the decorrelation energy. Of course, most such formulas focus on the shape between $0$ and $1$, since only that section is needed for the regular correlation energy. But any such approximate formula can be equally applied to $K_\mu$, yielding an approximation for the decorrelation energy. Peach et al. \cite{PMTT08} analyze various parametrizations for $W_\lambda$, and the same forms can be used as well to parametrize $K_\mu$, based on the similar shape of $W_\lambda$ and $K_\mu$ curves. In general, application to $K_\mu$  will yield a distinct approximation to the ground-state energy, with quite different properties.

To give just one example, one of the earliest sensible smooth parametrizations is the [1,1] Pade of Ernzerhof \cite{E96}:
\beq
W_\lambda = a \left( \frac{1+b\lambda}{1+c\lambda} \right).
\eeq
One can imagine applying it with inputs of e.g., $E\x$, $W'_0$ given by G\"{o}rling-Levy perturbation theory, and $W_\infty$ from the SC limit. It yields a sensible approximation to $W_\lambda$ in the 0 to 1 range, but because it was not designed with the strictly-correlated limit in mind, the formula itself is not immediately applicable to the decorrelation energy, since, e.g., $K_{\mu=0}$ vanishes. However, much more sensible is to make the same approximation directly for $K_\mu$ instead, if one is doing an SC calculation, i.e.,
\beq
K_\mu = \tilde{a} \left( \frac{1+\tilde{b}\mu}{1+\tilde{c}\mu} \right),
\eeq
whose inputs could be $K_0$, $K'_0$, and, e.g., a GGA for $K_{\mu=1}$. This is then a very different approximation from the same idea applied to the usual adiabatic connection formula.

On the other hand, there are several approximations designed to span the entire range of $\lambda$, the most famous being ISI (interaction strength interpolation) model \cite{SPKb00} developed by Seidl et al. This model uses the values and the derivatives of $W_\lambda$ at two limits, namely the high-density limit (KS system, $\lambda=0$) and the low-density limit (strictly-correlated system, $\lambda \to \infty$), to make an interpolation. Another approximation to $W_\lambda$ is developed in our previous work \cite{LB09}, which employs $W_0, W_\infty$ and $W_0'$ as inputs. We compare the approximate $K_\mu$'s obtained from the two models.

\begin{figure}[h]
\begin{center}
\includegraphics[width=3.5in]{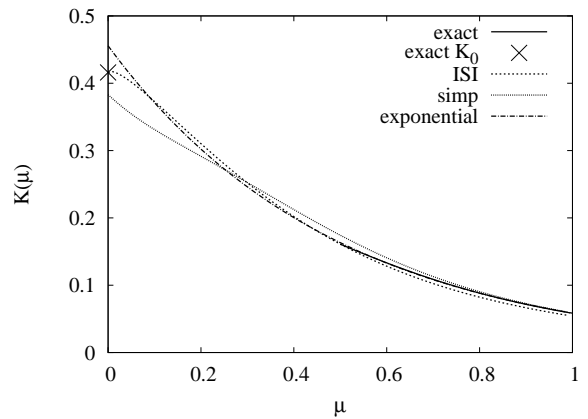}
\caption{Adiabatic connection curves for Hooke's atom. The exact curve (down to $\mu=0.5$) is taken from Ref. \cite{MTB03}.}
\label{hookeplot}
\end{center}
\end{figure}

Hooke's atom is a two-electron system (i.e., with Coulomb repulsion) in a spherical harmonic well \cite{T94}. Using the accurate values $W_0=-0.515, W_\infty=-0.743, W_0'=-0.101$ \cite{MTB03}, and $W_\infty'=0.208$ \cite{P09}, we find:
\begin{equation}
K_\mu^{\rm ISI} = -0.947\mu + 1.029 A\mu -\frac{0.336}{\mu B} +0.270 \mu \ln B,
\label{kmuisi4ho}
\end{equation}
where $A=\sqrt{1+0.653/ \mu^2}$ and $B=A-0.263$. With the same data substituted in $W^{\rm simp}$ \cite{LB09}, we find:
\begin{equation}
K_\mu^{\rm simp} = -\frac{0.228}{\alpha^4 \mu} (\alpha^3-\alpha^2+1)+1.287 \mu (\alpha-1),
\label{kmusimp4ho}
\end{equation}
where $\alpha=\sqrt{1+0.354/ \mu^2}$.
We plot the two forms of $K_\mu$ in Fig. \ref{hookeplot}. The exact curve (down to $\mu=0.5$) is taken from Ref. \cite{MTB03}. We compare three quantities in Table \ref{hooketable}. Although $K_\mu^{\rm ISI}$ contains a spurious $\mu \ln \mu$ term as $\mu \to 0$ \cite{S07, GVS09, LB09}, it nonetheless yields accurate results. The simple model, applied with the usual inputs, is less accurate pointwise, but integrates to an accurate value.

\begin{table}[h]
\caption{Comparison of several quantities for three approximations to $K_\mu$. Note: ISI uses $K_0$ as an input. The exact values are taken from Ref. \cite{MTB03}.}
\begin{center}
\begin{tabular}{c|c c c c}
\hline\hline
 & exact & ISI & simp & exponential\\
\hline
 $K_0$ & 0.416 & 0.416 & 0.383 & 0.456\\
 $K_1$ & 0.058 & 0.054 & 0.059 & 0.058\\
 $E\dc$ & 0.189 & 0.191 & 0.190 & 0.193\\
\hline\hline
\end{tabular}
\end{center}
\label{hooketable}
\end{table}

We can try the simple exponential parametrization Eq. (\ref{simpleunif}) for $K_\mu$ again for Hooke's atom. Because we do not know the value of $d/d\mu \ln K_\mu$ at $\mu=0$ exactly, instead we do an exponential fitting using the method of least squares, with the exact $K_\mu$ values (between $\mu=0.5$ and 1) taken from Ref. \cite{MTB03}. We plot $K_\mu$ vs. $\mu$ in Fig. \ref{hookeplot} and compare several quantities in Table \ref{hooketable}.

\subsection{H$_2$ Bond Dissociation}
Bond dissociation of the H$_2$ molecule produces a well-known dilemma in computational chemistry \cite{Koch, PSB95, B01, CMY08}. In the exact case, as the bond length $R \to \infty$, the hydrogen molecule should dissociate to two free hydrogen atoms, with the ground state always a singlet and spin-unpolarized. However, spin-restricted, e.g., restricted Hartree-Fock or restricted Kohn-Sham DFT, give the correct spin multiplicity, i.e. the wavefunction is an eigenfunction of $\hat{S}^2$, but produce an overestimated total energy, much higher than that of two free hydrogen atoms. Spin-unrestricted, e.g., unrestricted Hartree-Fock or unrestricted Kohn-Sham DFT, give a fairly good total energy, but the wavefunction is spin-contaminated, i.e., the deviation of $< \hat{S}^2 >$ from the exact value is significant. This is known as ``symmetry breaking'' in H$_2$ bond dissociation.

Fuchs et al. \cite{FNGB05} argued that DFT within RPA (random phase approximation) gives a correct picture of the H$_2$ bond dissociation within the spin-restricted KS scheme. They also gave highly-accurate adiabatic connection curves for ground-state H$_2$ at bond length $R=1.4$\AA\, and stretched  H$_2$ at bond length $R = 5$\AA. The curves were interpolated particularly between $\lambda=0$ and $1$, shown as the difference of the integrand, $\Delta W_\lambda$, between the stretched H$_2$ molecule and two free H atoms (Fig. 1 and 3 of Ref. \cite{FNGB05}). 

For $R=1.4$\AA\, and $R=5$\AA, if we use an interpolation (see Ref. 63 in Fuchs paper \cite{FNGB05}) to estimate $\Delta W_\lambda$, we find reasonable values $\Delta W_\infty=-7.00$ and $\Delta W_\infty=0.13$, respectively. Using Eq. (\ref{edcexc}), we find $\Delta E\dc=4.96$ and $\Delta E\dc=0.69$, respectively. We compare $\Delta E\dc$ and $\Delta E\c$ values in Table \ref{stretchedtable}. The comparison shows a physical example where the strictly-correlated system is a better starting point in the calculation.

\begin{table}[h]
\caption{Comparison of several quantities for stretched H$_2$ at different bond lengths. The values for $\Delta E\x$ and $\Delta E\xc$ are taken from Ref. \cite{FNGB05}. All values are in eV.}
\begin{center}
\begin{tabular}{c|c c c}
\hline\hline
bond length & $1.4$\AA & $5$\AA & $\infty$ \\
\hline
 $\Delta E\x$ & -0.98 & 5.85 & 8.5 \\
 $\Delta E\xc$ & -2.04 & 0.82 & 0.0 \\
 $\Delta W_\infty$ & -7.00 & 0.13 & 0.0 \\
 $\Delta E\c$ & -1.06 & -5.03 & -8.5 \\
 $\Delta E\dc$ & 4.96 & 0.69 & 0.0 \\
\hline\hline
\end{tabular}
\end{center}
\label{stretchedtable}
\end{table}

We can see that at the equilibrium bond length, $|\Delta E\c|$ is much smaller than $\Delta E\dc$, presumably making it easier to approximate the ground-state energy starting from the KS reference system.  This is typical at equilibrium. However for stretched bonds, $\Delta E\dc$ is much smaller than $|\Delta E\c|$, and so $\Delta E\dc$ instead may be better accurately approximated in the calculation and the strictly-correlated system should be a better reference. Molecules with strong static correlation, such as Cr$_2$ and O$_3$, might fall somewhere inbetween.

\section{Conclusion}

In this paper, we constructed an adiabatic connection formalism for the strictly-correlated system. We found that this adiabatic connection formula and curve can be well defined with respect to this new reference. Our formula connects the strictly-correlated system and the real system, by Eq. (\ref{updnac}). We also defined the quantity for this new integral ``decorrelation energy'' and related this with the usual KS adiabatic connection. We illustrated how the decorrelation energy behaves, using the uniform electron gas, Hooke's atom, and stretched H$_2$ as examples.

We emphasize again that a real application of this theory is only possible when the reference, i.e., the strictly-correlated system, can be routinely calculated. At present, one can calculate quantities such as $U_{\rm sc}$ exactly only for spherical symmetric systems \cite{SGS07}. However, nonempirical approximations to $E\xc$ of Kohn-Sham theory can be employed to estimate $W_\infty$ with useful accuracy \cite{PTSS04}. The computation of this quantity may become much easier in the future \cite{GSV09}. If this is so, then based on the properties discussed here, the strictly-correlated reference may be preferable in cases that are difficult for standard KS DFT calculations with standard approximations to $E\xc$. In fact, a recent work \cite{GSV09} independently shows progress using exactly the formalism discussed here and suggests approximations to $E\dc$. In any event, the advent of strictly-correlated calculations opens up a whole new alternative approach to DFT calculations of electronic structure, and only experience can show how and when this will prove more fruitful than the traditional (KS) scheme.

\section{Acknowledgement}
We thank John Perdew, Michael Seidl and Paola Gori-Giorgi for kind discussions. This work is supported by the National Science Foundation under No. CHE-0809859.

\bibliography{lib}

\end{document}